\documentclass[12pt]{iopart}

\usepackage{iopams}  

\usepackage{cite}
\usepackage{graphicx}%
\usepackage{hyperref}
\usepackage[dvipsnames]{xcolor}
\usepackage{ulem}

\begin{document}

\title[MXene Fe$_2$C as a promising candidate for the 2D XY ferromagnet]
{MXene Fe$_2$C as a promising candidate for the \\2D XY ferromagnet}

\author{E M Agapov$^{1}$, I A Kruglov$^{1,2}$, and A A Katanin$^{1,3,4}$}

\address{$^1$ Center for Photonics and 2D Materials, Moscow Institute of Physics and 
Technology, 141701 Dolgoprudny, Russia}
\address{$^2$ Dukhov Research Institute of Automatics (VNIIA), 127005 Moscow, Russian Federation}
\address{$^3$ Skolkovo Institute of Science and Technology, 121205 Moscow, Russia}
\address{$^4$ M. N. Mikheev Institute of Metal Physics, Russian Academy of Sciences, 620108 Yekaterinburg, Russia}

\ead{Andrey.Katanin@gmail.com}

\vspace{10pt}
\begin{indented}
\item[]July 2023
\end{indented}

\begin{abstract}
Monolayer Fe$_2$C is expected to possess strong electronic correlations, which can  significantly contribute to electronic and magnetic properties. In this work we consider electronic and magnetic properties of MXene Fe$_2$C within the DFT+DMFT approach.~We establish the existence of local magnetic moments  $\mu=3.2\mu_B$ in this compound, characterized by sufficiently long lifetime of $\tau\sim 350$~fs. We also calculate exchange interaction parameters accounting for electronic correlations using the recently developed approach for paramagnetic phase. We obtain the strongest exchange interaction $11$~meV between next nearest neighboring Fe atoms above (and below) the carbon plane, and the subleading interaction $6$~meV between the next to next nearest neighboring atoms across the carbon plane. The resulting dependence of the Berzinskii-Kosterlitz-Thouless (BKT) and Curie temperatures on magnetic anisotropy is obtained.
The BKT temperature for the pristine Fe$_2$C is $T_{\rm BKT}\simeq 290$~K, which makes this compound a good candidate for the two-dimensional ferromagnet with XY anisotropy. 
\end{abstract}

\section{Introduction}
The discovery of graphene in 2005 \cite{Graphene} gave a new impulse to the study of 2D materials. While graphene is non-magnetic, 
study of ferromagnetic materials is quite important,
since they provide realization of two-dimensional magnetism. According to the Mermin-Wagner theorem, the isotropic two-dimensional system do not possess long-range magnetic order at finite temperatures. Therefore, the magnetic phase transition in these materials is solely due to magnetic anisotropy. Strong magnetic correlations can nevertheless lead to the number of unusual properties, types of magnetic order, etc. 
This makes these materials also promising for possible application in new areas of science, such as, for example, spintronics. 

Starting from the suggestion of exfoliation of perovskite ferromagnetic three-dimensional compounds \cite{Kats2013}, the number of two-dimensional ferromagnetic materials were suggested, see, e.g. reviews \cite{2DFM1,2DFM2,2DFM3,2DFM4,2DFM5,2DFM6}. 
In particular, realization of MAX phases \cite{MAX} and then MXens opened up new perspectives. MAX phases are layered 3D materials of M$_{n+1}$AX$_n$ composition, where $n = 1$, $2$, or $3$, M is a transition metal, A is an element of the groups 13 or 14 of the periodic table, and X is carbon or nitrogen.
From such compounds it is sometimes possible to chemically etch (or peel off) the middle layer to get MXenes. These compounds have metallic properties, unique electrical 
capacitance, and show a high degree of intercalation. In view of these properties, MXenes have found application mainly in the field of electronics. This makes it is important to synthesize and study new MXenes, including those with the possibility of magnetic order. The representative in this class is Fe$_2$AC (A = Al, Si, Ge). 
The stability of the corresponding iron-based MAX phases was theoretically predicted \cite{Magn_MAX}. 
The presence of Fe allows to expect that this compound will have substantial magnetic moments, providing possibility of its strong magnetism.

The corresponding MXen Fe$_2$C is therefore a promising candidate for the class of two-dimensional magnets.
A density functional theory (DFT) study  \cite{YUE2017164} has shown from first principles that this material is expected to be thermodynamically stable and may possess ferromagnetic order with
a high Curie temperature. The presence of iron with its close to half filled d-orbitals makes the interelectronic interaction significant and causes electronic and magnetic correlations. 
In particular, they may yield appearance of local magnetic moments \cite{OurFe}, which are quite important for magnetic properties. Previous studies of elemental iron show, that the local magnetic moments are present due to Hund interaction and contribute essentially to thermodynamic and magnetic properties \cite{FeAdd1,OurFe,FeAdd2,FeAdd3,OurFe1}. 
The pristine Fe$_2$C is expected to posses XY (easy-plane) magnetic anisotropy \cite{YUE2017164}, and therefore it is expected to show the corresponding Berezinskii-Kosterlitz-Thouless (BKT) transition and finite magnetization in finite-size samples, similarly to that recently observed in CrCl$_3$ \cite{XY_FiniteSize_exp} and suggested for other Janus monolayers \cite{Janus1,Janus2}. 
Also, it was suggested that the anisotropy of Fe$_2$C can be changed to easy-axis by the external electric field \cite{LI2021126960}, deformation \cite{LOU2022169959}, and introducing layered geometry \cite{Sarmah_2023}.

In this work, we study electronic and magnetic properties of Fe$_2$C
using the combination of DFT with the dynamic mean field theory (DMFT). We show presence of local magnetic moments in this compound, and obtain, using the recently proposed technique \cite{MyJ}, exchange interactions between these local magnetic moments. On the basis of obtained exchange interactions we calculate the expected BKT and/or Curie temperature. The estimated temperature of BKT transition 
is essentially higher than in previously studied compounds. 
Based on the results of the study, it is possible to consider the applicability of the compound for spintronics or other areas using magnetic properties. Moreover, the proposed approach can be applied to study magnetic properties of other promising 2D materials and heterostructures.

\section{Computation details}
\begin{figure}[b]
\includegraphics{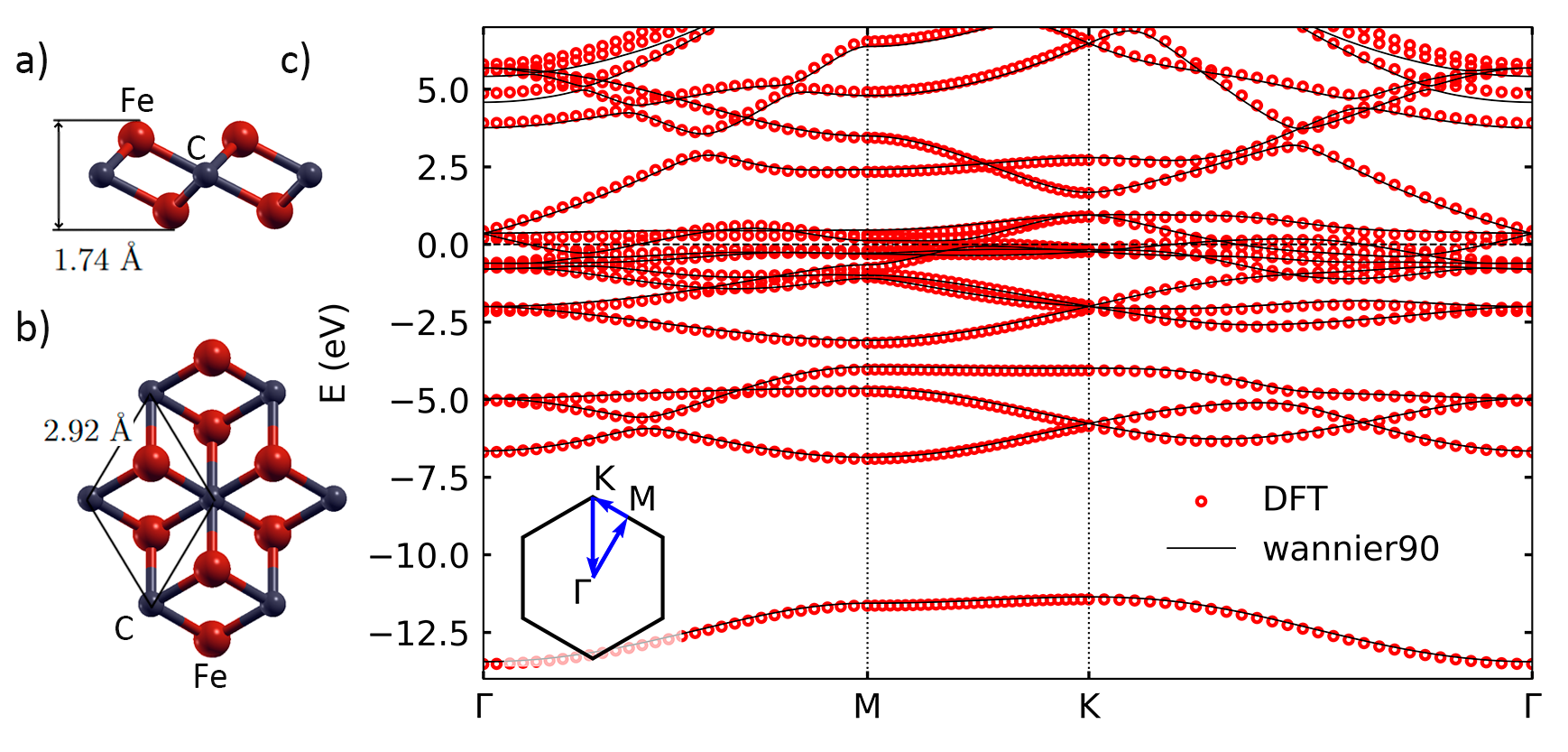} 
    \centering
    \caption{ Side (a) and top (b) view of Fe$_2$C (Fe atoms are shown by red, and C atoms by dark blue colour), and the band structure (c) of non-spin-polarized Fe$_2$C obtained from the DFT calculation (dots) and through the corresponding Wannier projection (lines). The Fermi level is at the zero energy. The inset shows the Brillouin zone with high-symmetry points.}
    \label{DFT_wannier}
\end{figure}

We consider the $1T$ crystal structure of Fe$_2$C with the space group P3$\bar{m}$1 (point group $D_{3d}$), see Fig. \ref{DFT_wannier}a,b. We perform the DFT calculations using Perdew-Burke-Ernzerhof (PBE) version of the generalized-gradient approximation (GGA) exchange-correlation functional realized in the Vienna ab initio simulation package (VASP) code \cite{VASP1, VASP2}. The projector augmented wave (PAW) potential \cite{PAW} is used with the plane-wave cutoff energy set as 500 eV. The Brillouin zone is sampled using a $30 \times 30 \times 1$ k-mesh for all calculations. 
The convergence threshold is set to be $10^{-6}$ eV per cell in energy and $0.01$~eV/\AA~for forces. In order to prevent interaction between layers through preriodic boundary conditions, the vacuum layer along c axis is set to $19$~\AA. During geometry optimisation, we relax the lattice vectors and atomic positions using the conjugate gradient method. After relaxation equilibrium lattice constant in nonspinpolarized calculations is found equal to $2.92$~\AA, 
Fe-Fe distance along the $z$ axis is equal to $1.74$~\AA, Fe-C distance is $1.68$~\AA. 


Electron interaction in iron-based materials, not fully considered by the DFT, is particularly important. To account of electronic correlations, we use the dynamical mean-field theory (DMFT) approach. After performing the DFT calculation, for being used as DMFT calculation input, the many-body Hamiltonian was constructed in a basis of 20 Wannier functions corresponding to $3d$ and $4p$ states of Fe and $2s,2p$ states of carbon using Wannier90 package \cite{Wannier90, Marzari1997MaximallyLG, Marzari_2012, MOSTOFI20142309}. Interpolated band structure is shown on figure \ref{DFT_wannier}. 

In DFT+DMFT calculations the $3d$ states of Fe were considered as correlated. There are 2 Fe atoms in the unit cell, one of them was treated within the impurity Anderson model, and the second one was taken into account using symmetry considerations. We consider the density-density interaction matrix, parameterized by Slater parameters $F^0$, $F^2$, and $F^4$, expressed through Hubbard $U$ and Hund $J_H$ interaction parameters according to 
${F^0\equiv U}$ and ${(F^2+F^4)/14 \equiv J_{\rm H}}$, $F^2/F^4\simeq 0.63$ (see Ref. \cite{u_and_j}). In the present work we take $U=4$~eV, $J_H=0.9$~eV, similarly to the previous studies of elemental iron \cite{OurFe1,OurFe}. We use a double-counting correction ${H}_{\rm DC} = M_{\rm DC} \sum_{im\sigma}  {n}_{im\sigma} $ in the around mean-field form~\cite{AMF},  $M_{\rm DC}=\langle {n}_{id} \rangle [U (2 n_{\rm orb} {-} 1) - J_{\rm H}  (n_{\rm orb} {-} 1)] / (2 n_{\rm orb})$, where
${n}_{id}$ is the {operator of the} number of $d$ electrons at the site $i$, $n_{\rm orb}=5$. To improve the applicability of the density-density form of the interaction, which neglects off-diagonal components of the local Green's functions with respect to the orbital indexes, for the considered low-symmetry two-dimensional compound, we perform the rotation of the basis of $d$-orbitals to diagonalize the local part of the Hamiltonian, corresponding to the crystal field.  

The DMFT calculations of the self-energies, non-uniform susceptibilities and exchange interactions were performed within the  hybridization expansion continous-time Quantum Monte Carlo (CT-QMC) method of the solution of impurity problem \cite{CT-QMC,CT-QMC1}, realized in the iQIST software package \cite{iQIST}.


\section{Results}
\subsection{Electronic and local magnetic properties}

\begin{figure}[t]
    \includegraphics{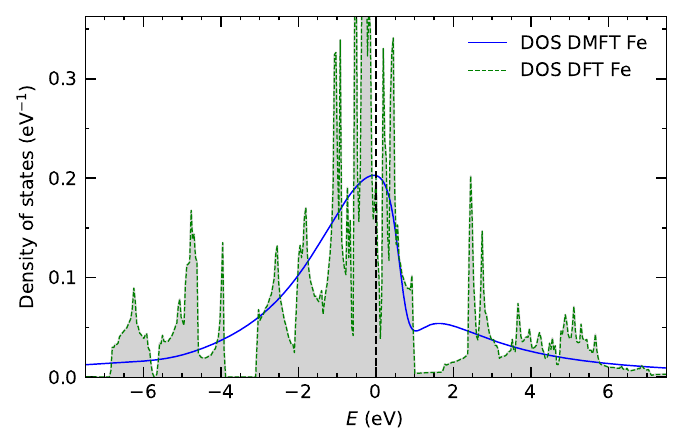}
    \centering
    \caption{The partial density of states of Fe in Fe$_2$C in DFT approach (the green dashed line) compared to the result of DFT+DMFT calculation at $\beta=1/T=10$~eV$^{-1}$ (solid blue line).}
    \centering
    \label{DMFT_DOS}
\end{figure}

\begin{figure}[t]
    \includegraphics{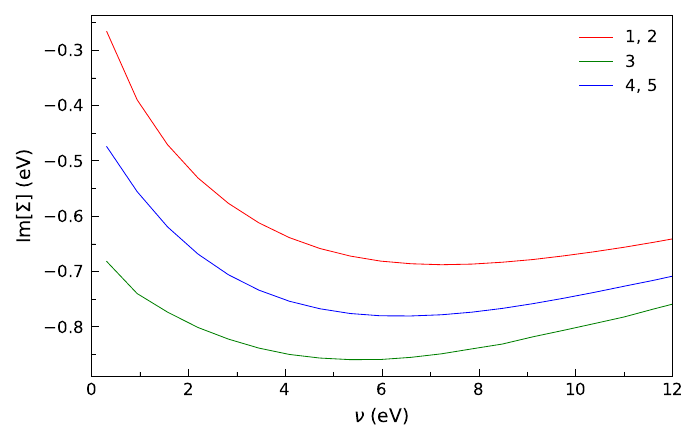}
    \centering
    \caption{Dependence of the imaginary part of self-energy on the imaginary frequency ${\rm Im}\Sigma(i\nu_n)$ at $\beta=10$~eV$^{-1}$. Numbers correspond to different local states after the basis rotation to diagonalize the crystal field: the states 4(1) and 2(5) are constructed mainly from the symmetric (antisymmetric) combination of the orbital states $d_{x^2-y^2}$, $d_{yz}$
and $d_{yz}$,  $d_{xy}$, and the state 3 is mainly from the orbital $d_{3z^2-r^2}$. The resulting states 1,2 and 4,5 correspond 
    to the $E_g$ irreducible representation of the point group $D_{3d}$ and 
    have pairwise degenerate self-energies and other local properties.}
    \centering
    \label{sgm_fr}
\end{figure}

Average filling of Fe $d$ orbitals in DFT+DMFT approach is 0.64 per one spin projection and sufficiently close to half filling, so that correlation effects are significant. The DFT+DMFT DOS at $\beta=1/T=10$~eV$^{-1}$ is shown in Fig. \ref{DMFT_DOS}, and it is suppressed at the Fermi level, similarly to the elemental iron \cite{OurFe}, showing effects of local correlations, related to the presence of local magnetic moments. The suppression of the spectral function originates from large damping of electronic states, as can be seen  from the imaginary part of the self-energy at the imaginary frequency axis, shown on the figure \ref{sgm_fr}. The largest damping ($\sim 0.7$~eV at $\beta=10$~eV$^{-1}$) corresponds to the $d_{3z^2-r^2}$ states, other states have smaller damping, $\sim 0.45$~eV and $\sim 0.35$~eV, which is still larger than temperature.

\begin{figure}[t]
    \includegraphics{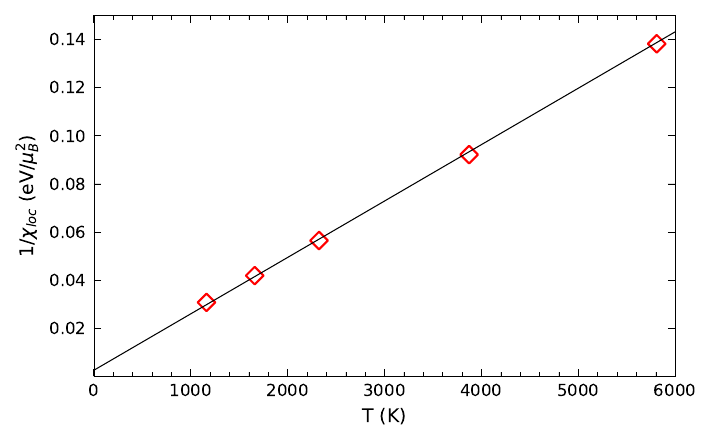}
    \centering
    \caption{Temperature dependence of the inverse local static spin susceptibility on temperature.}
    \centering
    \label{chi_S_T}
\end{figure}

To show the existence of local magnetic moments and estimate their size, we consider the static limit $\chi_{\rm loc}=\chi_{\rm loc}(\omega_n=0)$ of the dynamic local  spin susceptibility $\chi_{{\rm loc}}\left(i \omega_n\right)=(g\mu_B)^2\int_0^\beta \exp \left(i \omega_n \tau\right)\left\langle S^z_i(\tau) S^z_i(0)\right\rangle$, 
where $S^z_i=\sum_m S^z
_{i,m}$ is the local spin operator at site $i$, 
$S^z_{i,m}=(1 / 2) \sum_{\sigma \sigma^{\prime}} c_{i m \sigma}^{+} \sigma^z_{\sigma \sigma^{\prime}} c_{i m \sigma^{\prime}}$, $c_{i m \sigma}^{+}$, $c_{i m \sigma}$ - electron creation and annihilation operators at site $i$, orbital $m$, $\sigma^z$ is the Pauli matrix, $\beta$ is the inverse temperature, and $\omega_n$ are bosonic Matsubara frequencies. The temperature dependence of the inverse static susceptibility $\chi_{\rm loc}$ is shown in figure \ref{chi_S_T}. It obeys the Curie-Weiss law $\chi_{\rm loc} = {\mu_{\rm loc}^2}/({3(T - \theta)})$, reflecting existence of local magnetic moments, where the negative Weiss temperature $\theta = -\sqrt{2}T_K$ is related to the Kondo temperature $T_K$, see Refs. \cite{Sangiovanni,MyComment} and references therein. The obtained $T_K$ is quite low, showing well formed local magnetic moments. 
From the slope of the obtained dependence we find $\mu_{\rm loc}^2 \simeq 11 \mu_B^2$, corresponding to the effective spin $p=1.2$ defined by $\mu_{\rm loc}^2=(g\mu_B)^2 p(p+1)$.



\begin{figure}[b]
    \includegraphics{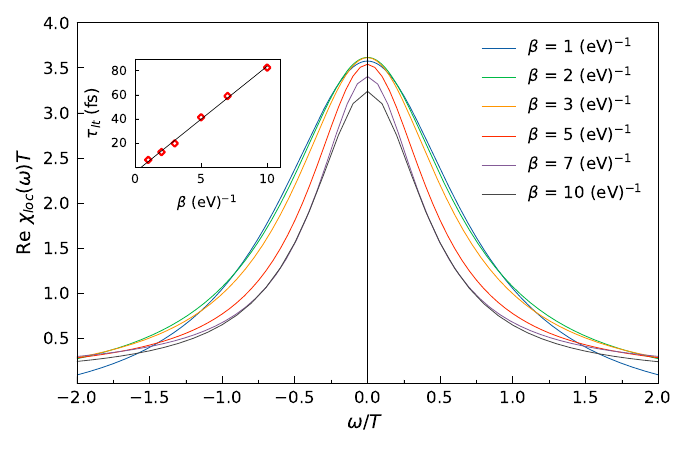}
    \centering
    \caption{ The main plot shows the real parts of the local dynamic susceptibilities obtained from DMFT at different temperatures. The embedded panel shows the dependence of the lifetime of local moments on temperature}
    \centering
    \label{chi_w}
\end{figure}

To obtain magnetic moments lifetime we further study  the local dynamic susceptibility. To obtain the susceptibility at real frequencies, we perform the analytical continuation of $\chi_{\mathrm{loc}}\left(i \omega_n\right)$ by using the Pade approximants. The dependence of its real part on frequency at various temperatures is shown on main panel of figure \ref{chi_w}.  Magnetic moments inverse lifetimes $\tau_{\rm lt}$ were obtained as half-width of the peaks of local dynamic susceptibility \cite{ToschiPnic,OurGammaFe,ToschiTime
}. They scale linearly with temperature, with $h/\tau_{\rm lt}\sim 0.5T$ ($h$ is the Planck's constant). The extrapolation to room temperature (300 K) gives lifetime $350$~fs.

\subsection{Nonlocal magnetic properties}
\begin{figure}[b]
    \includegraphics{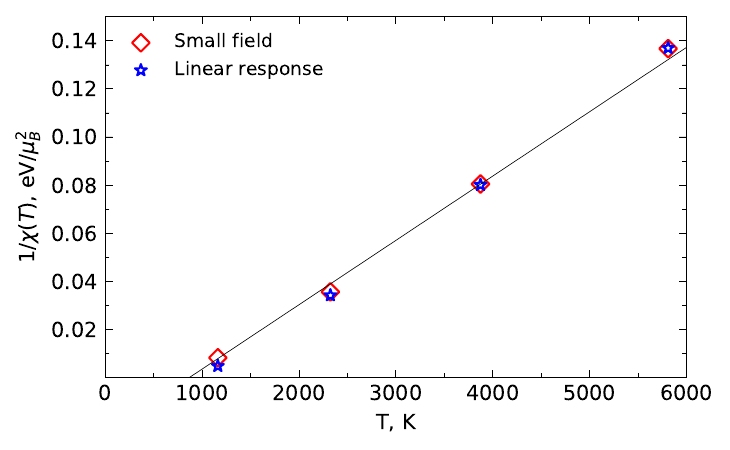}
    \centering
    \caption{Dependence of the uniform magnetic susceptibility on temperature. The red diamonds indicate the points obtained from the calculation in the presence of small magnetic field and the blue stars indicate the points obtained from the linear response theory.}
    \centering
    \label{chi_T}
\end{figure}

To obtain the DMFT Curie temperature the uniform magnetic susceptibility $\chi(T) = {M(T)}/{H}$ was calculated with constant magnetic field $H = 0.0025$ eV/$\mu_B$. The magnetic susceptibility was also obtained from calculations in the paramagnetic state, as a limit ${\mathbf q}\rightarrow 0$ of the non-uniform susceptibility $\chi_{\mathbf q}$, expressed using the linear response theory via dynamic vertices, see below and  Refs. \cite{MyJ, RevModPhys.90.025003}. Both dependencies are shown on figure \ref{chi_T} and agree with each other. 
By extrapolating the obtained dependence to zero, we obtain the  temperature of magnetic transition $T^{\rm DMFT}_C = 865$ K.
Due to the two-dimensionality, the inverse susceptibility in the absence of anisotropy should in fact exponentially approach zero at zero temperature \cite{2D1,2D2},
and in the presence of anisotropy it should approach zero at the Curie temperature, which is lower than the obtained DMFT Curie temperature. The latter corresponds to the transition to the regime of strong magnetic correlations. Therefore, the actual Curie temperature differs from the obtained DMFT value, and estimated below in Sect. \ref{SectCurie}.
According to the Curie-Weiss law $\chi = {\mu^2}/({3(T - T^{\rm DMFT}_C)})$, we find the local magnetic moment from the uniform susceptibility $\mu^2 = 9.66 \mu_B^2$. Relating this to the local spin via $\mu^2 =  \mu_B^2 g^2 S(S+1)$ we obtain $S \approx 1$. The fact that the spin $S$  does not depend appreciably on temperature and close to the integer value is inherent for local magnetic moments, which are present in the Fe-based materials.

\begin{figure}[t]
    \includegraphics{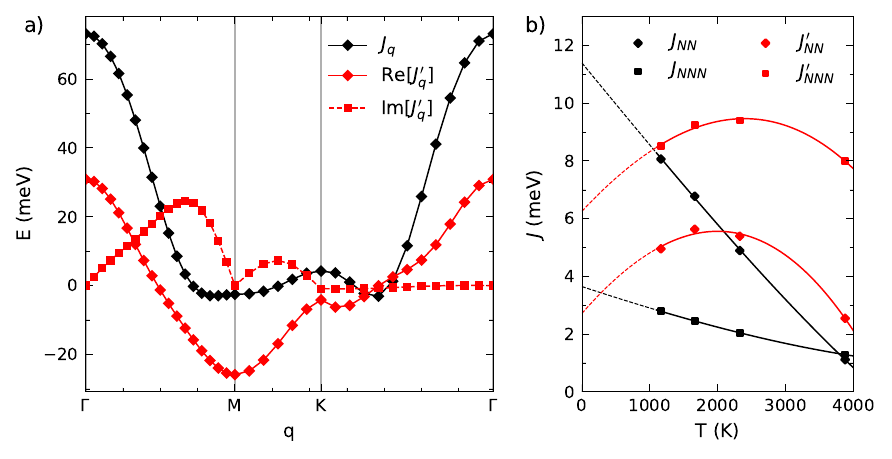}
    \centering
    \caption{a) Momentum dependence of the exchange interaction along the high symmetry directions in Fe$_2$C at $T = 1160$~K ($\beta=10$~eV$^{-1}$) between Fe atoms of the same layer ($J_{\mathbf q}$, black rhombs) and different layers across the carbon plane ($J'_{\mathbf q}$, red rhombs (real part) and squares (imaginary part)). b) Temperature dependence of the exchange interaction parameters. Dashed lines show the extrapolation to the low-temperature region.}
    \centering
    \label{J_q}
\end{figure}

To estimate the strength of exchange interaction between the local magnetic moments within the Heisenberg model, the random phase approximation (RPA) for spin degrees of freedom was used, see \cite{MyJ} and references therein. In particular, we generalize the approach of Ref. \cite{MyJ} considering the orbital-summed susceptibilities, to the two atoms per unit cell, $\hat{J}_{\mathbf{q}}={\hat{\chi}}_{\mathrm{loc}}^{-1}-{\hat{\chi}}_{\mathbf{q}}^{-1}$ where hats stand for matrices $2\times2$ in the atom number in the unit cell, $\chi^{rr'}_{\mathbf{q}}=-\langle\langle S_{\mathbf{q},r}^z \mid S_{-\mathbf{q},r'}^z\rangle\rangle_{\omega=0}$ is the matrix of the non-local static longitudinal susceptibilities,
$\chi^{rr'}_{\mathrm{loc}}=-\left\langle\left\langle S_i^z \mid S_i^z\right\rangle\right\rangle_{\omega=0}\delta_{rr'}$ is the diagonal matrix of local spin susceptibilities. 
Here $S^z_{\mathbf{q},r}=\sum_{{\mathbf R}_{i_u}} \exp(i {\mathbf q}{\mathbf R}_{i_u})S^z_{\mathbf{R}_{i_u},r}$ is the Fourier component of the local spin operators $S^z_i$ with respect to the coordinates ${\mathbf R}_{i_u}$ of the unit cell of atom $i=(i_u,r)$, $i_u$ is the index of the unit cell, $r,r'=1,2$ are the indexes of atoms in the unit cell.
Accordingly, we consider the interaction between the atoms of the same Fe plane (above or below carbon atoms plane), $J^{11}_{\mathbf q}=J^{22}_{\mathbf q} = J_{\mathbf q}$ and the interplane interaction across the carbon plane, $J^{12}_{\mathbf q}=(J^{21}_{\mathbf q})^* = J_{\mathbf q}'$. 

\begin{figure}[t]
    \includegraphics{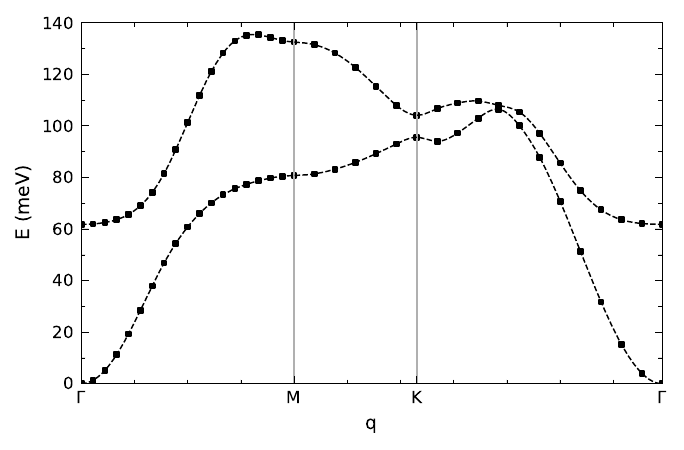}
    \centering
    \caption{Magnon dispersion in Fe$_2$C at $T = 1160$ K ($\beta=10$~eV$^{-1}$) along the high symmetry directions. }
    \centering
    \label{magn_disp}
\end{figure}

The nonlocal susceptibilities $\hat{\chi}_{\mathbf{q}}^{-1}$ are extracted from the  summation of ladder diagrams, containing particle-hole bubbles connected by the irreducible local vertex in the spin channel; the latter is extracted from the DFT+DMFT analysis, see Refs. \cite{RevModPhys.90.025003,MyJ}. The resulting momentum dependencies of $J_{\mathbf q}$ and $J'_{\mathbf q}$ are shown in Fig. \ref{J_q}a. Both interactions are maximal at $q=\Gamma=0$, showing strong tendency to ferromagnetic order. The temperature dependence of the exchange integrals is depicted on the right panel of figure \ref{J_q}. The results show that the exchange interaction decreases with increasing temperature. 
The contribution to the interaction falls off rapidly for atoms at large distances. At 
low temperatures nearest (next nearest) neighbours contribute $11.2$~meV ($3.8$~meV) to the exchange interaction among the atoms within the same Fe plane, located at the distance 2.92 $\AA$ (5.05 $\AA$) and $2.8$~meV ($6.1$~meV) to the exchange interaction between the Fe atoms across carbon plane, located at the distance 2.41 $\AA$ (3.79 $\AA$).   

\begin{figure}[b]
    \includegraphics{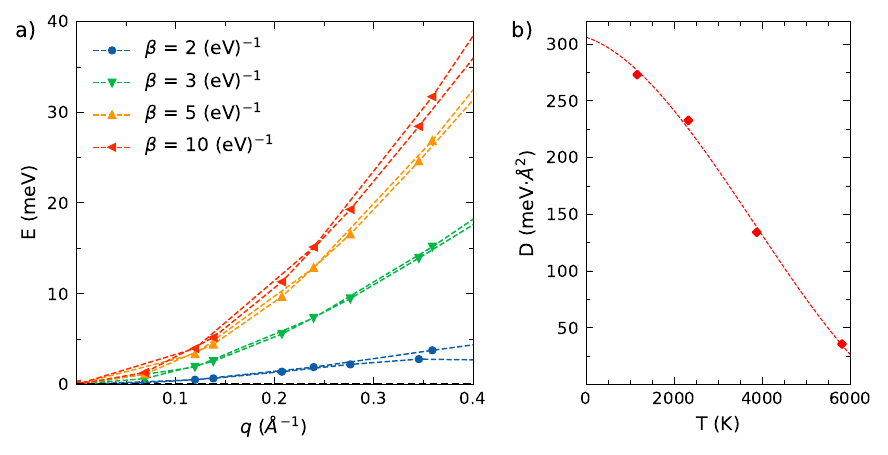}
    \centering
    \caption{ a) Magnon dispersion at small momenta and various temperatures. b) Temperature dependence of the spin-wave stiffness $D$.  Dashed line shows the extrapolation.}
    \centering
    \label{q_magn_disp}
\end{figure}

We retrieve magnon dispersion from exchange interaction momentum dependence along the symmetry directions. Representing exchange interaction  
$H= -({1}/{2}) \sum_{\mathbf{q},rr'} J^{rr'}_{\mathbf{-q}} \mathbf{S}_{\mathbf{q},r} \mathbf{S}_{-\mathbf{q},r'}$ in previous notations we consider lowest order (quadratic) terms in Dyson-Maleev or Holstein-Primakoff representation and diagonalize the corresponding Hamiltonian \begin{equation}
    \hat{H} = 
    S\left(\begin{array}{cc}
        \hat{a}_{\mathbf{q},1}^{+} \\
        \hat{a}_{\mathbf{q},2}^{2^+} 
       \end{array}\right)
    \left(\begin{array}{cc}
        J_0^{11} + 
        J_0^{12}
- J_{-\mathbf{q}}^{11} & - J_{-\mathbf{q}}^{12} \\
        - J_{-\mathbf{q}}^{21} & J_0^{22} + 
        J_0^{21} 
        - J_{-\mathbf{q}}^{22} \\
    \end{array}\right)
    \left(\begin{array}{cc}
        \hat{a}_{\mathbf{q},1} \\
        \hat{a}_{\mathbf{q},2} 
    \end{array}\right)
\end{equation}
where $\hat{a}_{\mathbf{q},r}^{+}, \hat{a}_{\mathbf{q},r}$ are the operators of creation and annihilation of magnon at the site $r$ of the unit cell.
The corresponding dispersion is shown in the figure \ref{magn_disp}. The obtained dispersion is positive, showing once more the stability of ferromagnetic order. From the small momentum part of the dispersion (see figure \ref{q_magn_disp}(a))
we obtain the temperature dependence of the spin-wave  stiffness $D$ (figure \ref{q_magn_disp}(b)). In the low temperature limit we obtain $D\simeq 300$~meV$\cdot$\AA$^2$, which is comparable to pure iron, see Refs. \cite{Fe1,Fe2,Fe3}.

\subsection{Curie temperature}
\label{SectCurie}
According to the Mermin-Wagner theorem, in an isotropic two-dimensional system, there cannot be long-range magnetic ordering at finite temperatures, that is, 
the Curie temperature $T_C = 0$.
Thus, the finite Curie temperature appears solely due to the  magnetic anisotropy energy ({{\textrm{MAE}}}), not considered in the performed calculations. 
It was found in \cite{YUE2017164} that the \textmd{MAE} of pure Fe$_2$C is negative (i.e. easy plane), and equals to $-2\cdot 10^{-5}$~eV per Fe atom. Therefore, pure Fe$_2$C is expected to show Berezinskii-Kosterlitz-Thouless transition. The corresponding temperature of BKT transition can be obtained from the equation \cite{MyXY}
\begin{equation}
     T_{\rm BKT} = \frac{4 \pi \rho_s}{\ln{({T_{\rm BKT}}}/{ |\Delta|})+4\ln ({4\pi \rho_s/T_{\rm BKT})}}
     \label{EqTBKT}
\end{equation}
\begin{figure}[t]
\includegraphics[width=0.65\linewidth]{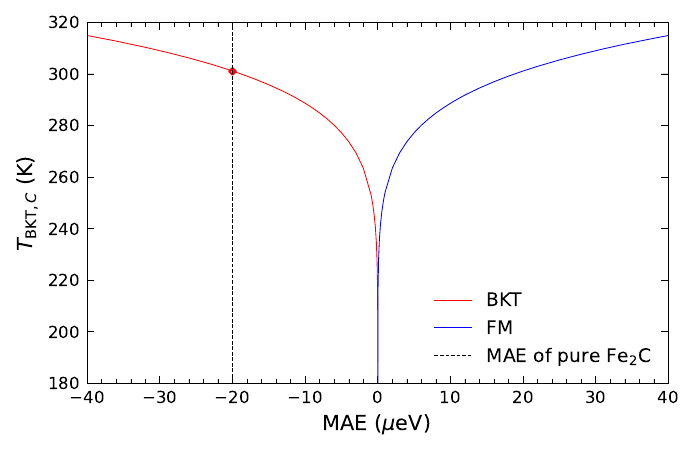}
    \centering
    \caption{The MAE dependence of $T_{{\rm BKT},C}$. The MAE~$<0$ part corresponds to the BKT phase transition, which becomes ferromagnetic for finite size systems or in the presence of magnetic substrate, the MAE~$>0$ part corresponds to the ferromagnetic phase transition. The dashed line shows the anisotropy of pure Fe$_2$C per atom from Ref. \cite{YUE2017164}.}
    \centering
    \label{T_MAE}
\end{figure}
where $\Delta(T) = (2S - 1)/S^2 \cdot {{{\textrm{MAE}}}}$, $\rho_s = DS/A$ is the spin stiffness, $A$ is the cell area. We note that the finite size easy plane magnets are expected to show finite magnetization with the Curie temperature \cite{XY_FiniteSize}
\begin{equation}
T_{\rm C}\simeq T_{\rm BKT}+\frac{C}{\ln^2 {L}} 
\label{TCXY}
\end{equation}
($C$ is some constant, $L$ is the linear system size), obtained from the condition of the equality of correlation length above BKT temperature to the system size. This was also recently experimentally observed in CrCl$_3$ monolayer \cite{XY_FiniteSize_exp}.
Also, small coupling to the magnetic substrate acts as an interlayer coupling and yields an onset of long-range ferromagnetc order with the Curie temperature $T_C$, determined by the Eq. (\ref{TCXY}) with $L=J/J'$, where $J'$ is the interlayer coupling, see Refs. \cite{XY_Layered,MyXY,XY_FiniteSize}. Apart from that, several possible ways to change the sign of magnetic anisotropy of Fe$_2$C was suggested in the literature, including introduction of the external electric field \cite{LI2021126960}, deformation \cite{LOU2022169959}, and layered geometry \cite{Sarmah_2023}. In case of positive (easy-axis) anisotropy $\Delta>0$, the same equation (\ref{EqTBKT}) is applicable with the replacement $T_{\rm BKT}\rightarrow T_C$, see Ref. \cite{PhysRevB.60.1082}.

Using $A = 7.24$~\AA$^2$ and the extrapolated spin wave stiffness $D=287$ meV$\cdot \AA^2$, we obtain $\rho_s = 38.9$ meV. The respective dependence $T_{{\rm BKT},C}(\textrm{MAE})$ is plot on Fig. \ref{T_MAE}. For the anisotropy of pure Fe$_2$C we obtain a rather high value $T_{\rm BKT}\simeq 290$~K.

\section{Conclusions}
In this study, the magnetic properties of the Fe$_2$C were explored in DFT+DMFT framework.~From DFT+DMFT approach we have  obtained the density of states and electronic self-energies accounting for electronic correlations.~From the local spin susceptibility we have shown that Fe$_2$C possesses a local magnetic moment of 3.2 $\mu_B$ with a lifetime of $\simeq 350$ fs. The uniform  magnetic susceptibility follows the Curie-Weiss law confirming existence of local magnetic moments. The obtained DFT+DMFT Curie temperature $T^{\rm DMFT}_C=865$~K is rather large, but it only marks the onset of strong ferromagnetic correlations.

The exchange interactions, extracted from the inverse nonlocal susceptibilities, show maximum at the $\Gamma$ point of the Brillouin zone; the obtained magnon dispersions are positive showing stability of the ferromagnetic order. The obtained nearest- and next-nearest neighbor exchange interactions are in the range of $2$-$11$~meV. 

Using the obtained exchange interactions, we have estimated the spin-wave stiffness at low temperatures $D\simeq 300$~K and estimated the anisotropy dependence of Berezinskii-Kosterlitz-Thouless $T_{{\rm BKT}}$ (for easy plane anisotropy) or Curie temperature (for easy-axis anisotropy). For the easy plane anisotropy of pristine Fe$_2$C we obtain $T_{{\rm BKT}}\simeq 290$ K, which makes this compound a good candidate for the two-dimensional ferromagnet with XY anisotropy. 

The experimental realization of this material seems therefore quite interesting. Apart from that, the methods developed in this study can be furthermore applied to the other perspective two-dimensional compounds.

\section*{Acknowledgements}
E. M. A. and I. A. K. (performing DFT+DMFT calculations) acknowledge the financial support from the Ministry of Science and Higher Education of the Russian Federation (Agreement No. 075-15-2021-606). The work of A. A. K. (developing DFT+DMFT approach for calculation of local and non-local quantities, as well as the exchange interactions) was supported by RSF grant 19-72-30043-P. 

\vspace{1cm}
\bibliographystyle{iopart-num}
\bibliography{bibliography.bib}

\end{document}